\documentclass[sigconf,nonacm]{acmart}

\usepackage{url}
\usepackage{graphicx}
\usepackage{hyperref}

\AtBeginDocument{%
  }

\setcopyright{acmlicensed}
\copyrightyear{2025}
\acmYear{2025}
\acmDOI{XXXXXXX.XXXXXXX}
\acmConference[MSR 2026]{MSR '26: Proceedings of the 23rd International Conference on Mining Software Repositories}{April 2026}{Rio de Janeiro, Brazil}
\acmISBN{978-1-4503-XXXX-X/2018/06}

\newcommand{\summarybox}[2]{%
\begin{center}
\fbox{%
    \begin{minipage}{0.95\linewidth}
        \textbf{Summary of #1:} #2
    \end{minipage}
}
\end{center}
}

\begin{document}

\title{How Do Agentic AI Systems Address Performance Optimizations? A BERTopic-Based Analysis of Pull Requests}

\author{Md Nahidul Islam Opu}
\affiliation{%
  \institution{SQM Research Lab\\ Computer Science\\ University of Manitoba}
  \country{Winnipeg, Canada}
}

\author{Shahidul Islam}
\affiliation{%
  \institution{SQM Research Lab\\ Computer Science\\ University of Manitoba}
  \country{Winnipeg, Canada}
}

\author{Muhammad Asaduzzaman}
\affiliation{%
  \institution{Computer Science\\ University of Windsor}
  \country{Windsor, Canada}
}

\author{Shaiful Chowdhury}
\affiliation{%
  \institution{SQM Research Lab\\ Computer Science\\ University of Manitoba}
  \country{Winnipeg, Canada}
}


\begin{abstract}

LLM-based software engineering is influencing modern software development. In addition to correctness, prior studies have also examined the performance of software artifacts generated by AI agents. However, it is unclear how exactly the agentic AI systems address performance concerns in practice. In this paper, we present an empirical study of performance-related pull requests generated by AI agents. Using LLM-assisted detection and BERTopic-based topic modeling, we identified 52 performance-related topics grouped into 10 higher-level categories. Our results show that AI agents apply performance optimizations across diverse layers of the software stack and that the type of optimization significantly affects pull request acceptance rates and review times. We also found that performance optimization by AI agents primarily occurs during the development phase, with less focus on the maintenance phase. Our findings provide empirical evidence that can support the evaluation and improvement of agentic AI systems with respect to their performance optimization behaviors and review outcomes.


\end{abstract}





\maketitle

\section{Introduction}

LLM-based software engineering is rapidly reshaping development practices, with autonomous agents now implementing features, fixing bugs, and submitting pull requests (PRs) to software repositories in-the-wild~\cite{li2025rise}. Beyond functional correctness, the performance of AI-generated code is critical due to its impact on scalability, efficiency, and user experience~\cite{chen2021evaluating,peng2025perfcodegen}. As a result, prior work has evaluated the performance of LLM-generated code against human-written code~\cite{chen2021evaluating,wang2024performance,watanabe_use_2025}, reporting mixed outcomes: AI can match or exceed human performance in some settings~\cite{coignion2024performance,wang2024performance,peng2025perfcodegen}, but shows notable inefficiencies in others~\cite{abbassi2025taxonomy,diehl2025llm,yi_experimental_2025}.

However, existing studies largely rely on outcome-based evaluations conducted in controlled settings~\cite{abbassi2025taxonomy,diehl2025llm}, providing limited insight into how agentic AI systems address performance concerns within real-world development workflows. In particular, it remains unclear which types of performance issues AI agents target and at which layers of the software stack these optimizations are applied. Moreover, little is known about the Software Development Life Cycle (SDLC) activities in which AI agents introduce performance optimizations, or whether different optimization types influence pull request acceptance rates and review times. Addressing these gaps is essential for enabling deeper empirical analyses that can inform the design and deployment of performance-aware LLM-based agents~\cite{peng2025perfcodegen,yi_experimental_2025}. Motivated by these gaps, we answer the following three research questions by analyzing a set of AI-generated performance-related pull requests as provided by the publicly available AIDev dataset~\cite{li2025rise}.

\textbf{RQ1: What categories of performance optimizations do agentic AIs address?}
Analyzing optimization categories reveals whether agents focus on localized changes or span the software stack, informing future model training and tooling. We identified 52 performance topics grouped into 10 categories, ranging from low-level optimizations to UI- and AI-related concerns, showing that agentic AIs operate across multiple stack layers.

\textbf{RQ2: How do PR rejection rates and review times vary across categories?}
Examining PR outcomes highlights areas where agentic AIs are more readily trusted versus those needing improvement. We observed an overall rejection rate of 36.5\% in performance-related PRs, which is significantly higher than in non-performance PRs (22.7\%). Also, rejection rates and review latency are largely driven by the type of performance concern, with agents performing worst on \textit{UI}, \textit{AI}, and \textit{Analytics} PRs.

\textbf{RQ3: What SDLC activities are associated with AI-generated performance-related PRs?}
We found that performance-related PRs are more associated with feature implementation (i.e., development activities) and comparatively less associated with bug fixes, refactoring, and testing. This may indicate a potential gap in AI-assisted performance optimization during the maintenance phase.

To enable replication, we share our data and code publicly.\footnote{\url{https://github.com/SQMLab/LLM-performance}}

\begin{figure*}[ht]
    \centering
    \includegraphics[width=1\linewidth, keepaspectratio]{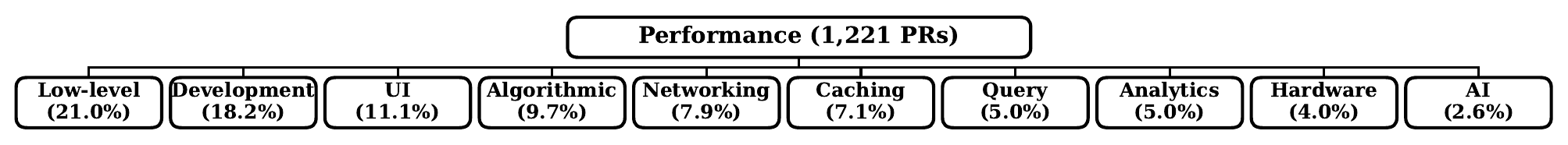}
    \caption{Performance-related PRs grouped into 10 categories.}
    \Description{Performance-related PRs grouped into 10 categories.}
    \label{fig:pr_tree}
\end{figure*}
\section{Methodology}


We used the AIDev-POP subset of version v3 of the AIDev dataset~\cite{li2025rise}, obtained from Zenodo~\cite{li_2025_16919272}, which comprises 33{,}596 agentic pull requests. We focus only on this subset as it captures higher-quality, real-world projects and has the potential to avoid toy repositories~\cite{kalliamvakou2014promises}. Also, it provides the metadata required for RQ2 and RQ3, including PR acceptance status, review timing, and SDLC activities.

\textbf{Identifying Performance-Related PRs.}
AIDev dataset provides LLM-generated PR type annotations for each PR, but only 340 PRs are labeled as \textit{perf}, substantially under-representing performance-related changes. We found that many PRs labeled as bug fixes or features also address performance concerns. For instance, the PR titled \emph{'Fix Gantt performance by removing useMouse from every column'}\footnote{\url{https://github.com/shadcnblocks/kibo/pull/166}} is labeled as a \textit{fix} despite explicitly targeting performance. To address this limitation, we aimed to identify all PRs that express performance concerns, even if they are associated with other activities, such as bug fixes. 

Following prior works~\cite{das2016quantitative,nistor2013discovering,selakovic2016performance,zaman2012qualitative}, we initially applied keyword-based filtering using 23 performance-related terms (e.g., ``performance, ``optimize'', ``latency''). Although this process yielded 10{,}894 PRs, manual inspection of 100 randomly sampled PRs by the first author---a graduate student with 2+ years of industry experience---shows a 92\% false positives, making this approach unsuitable.

We therefore adopted an LLM-based detection approach and reformulated the task as a binary classification problem (performance vs.\ non-performance). Using the open-source \textit{gpt-oss-20B} model~\cite{agarwal2025gpt} in a zero-shot setting~\cite{kojima2022large}, we classified PRs based on their intent. The prompt instructs the LLM to label a PR as performance-related only when performance is expressed as a concern. To guide the LLM, the prompt also included the set of performance-related keywords; the full prompt is provided in the replication package. 

This process identified 1{,}160 performance-related PRs across 447 repositories, generated by five different agents. Manual inspection of 200 randomly sampled PRs independently by first two authors, graduate students with 2+ and 10+ years of industry experience as software engineers, revealed a 7.5\% false positive rate, defined as cases where at least one author labeled a PR as NonPerformance. Observed agreement was 97.5\%, with high inter-rater reliability as measured by Cohen’s kappa (0.79)~\cite{cohen1960coefficient} and Gwet’s AC1 (0.97)~\cite{gwet2008computing}. However, since we applied BERTopic with Hierarchical Density-Based Spatial Clustering of Applications with Noise (HDBSCAN), which detects outliers~\cite{McInnes2017}, we did not have to filter these cases, thus avoiding another step of laborious manual annotation. We also found that our new dataset missed 61 PRs from the 340 \textit{perf}-labeled PRs in AIDev. To enlarge the dataset size, we added those 61 samples, yielding a final dataset of 1{,}221 performance PRs, substantially larger than the original subset. 


\textbf{Topic Modeling with BERTopic.}
We applied the BERTopic~\cite{grootendorst2022bertopic} due to its superior performance over traditional methods~\cite{egger2022topic,el2024comparative,gan2023experimental} and identify topics in the filtered PRs. Combining the title and body of a PR, we generated embeddings using \textit{Qwen3-Embedding-8B}, which currently ranks top for clustering tasks on the MTEB benchmark~\footnote{\url{http://mteb-leaderboard.hf.space/?benchmark_name=MTEB\%28Multilingual\%2C+v2\%29}
, last accessed: December 20, 2025.}.

We reduced embeddings using Uniform Manifold Approximation and Projection (UMAP)~\cite{mcinnes2018umap} and performed topic clustering with HDBSCAN~\cite{McInnes2017}. To refine topic representations, we followed the standard BERTopic pipeline~\cite{grootendorst2022bertopic} and applied class-based TF-IDF, CountVectorizer-based term filtering, and a hybrid strategy combining part-of-speech extraction with Maximal Marginal Relevance. 


Given the large BERTopic parameter space and multiple evaluation criteria, we conducted a constrained grid search evaluating topic coherence, which assesses semantic interpretability~\cite{roder2015exploring}, and the silhouette score, which measures cluster separation and compactness~\cite{rousseeuw1987silhouettes}. We selected the configuration that maximizes both metrics (coherence score of 0.47 and silhouette score of 0.57), yielding UMAP parameters $n\_components=20$ and $n\_neighbors=3$, and HDBSCAN parameters $min\_cluster\_size=10$ and $min\_samples=1$. This configuration produced 52 topics and 101 outliers, consistent with topic counts reported in some prior SE studies~\cite{bangash2019developers,uddin2021empirical}.


To assign semantic labels, the first two authors independently labeled all topics based on topic keywords and randomly sampled PRs. The Cohen's Kappa score for inter-rate agreement was 0.92, and the four labeling disagreements were resolved through discussion. Detailed topic labels are included in the replication package.

\textbf{Statistical Tests.} In this paper, we use the non-parametric Wilcoxon rank-sum test, Cliff's delta effect size, and Kendall's $\tau$ correlation coefficients as these statistical tests are commonly used in SE research~\cite{Inozemtseva:2014, greenscaler:2019, Bangash}.

\section{Approach, Analysis, and Results}
This section presents the RQ-specific methodology and results.

\subsection{RQ1: Topics and Categories}


BERTopic produced 52 distinct topics from the identified performance-related PRs. After topic labeling, we merged semantically related topics into higher-level categories. The first two authors collaboratively derived these categories through iterative discussion. This process resulted in 10 overarching categories, shown in Figure~\ref{fig:pr_tree}.


The resulting categories span multiple layers of the software stack, ranging from \textit{Low-level} optimizations (e.g., compiler and transpiler improvements) to higher-level concerns such as \textit{UI} performance and \textit{AI}-specific optimizations (e.g., token usage and ChatAPI efficiency). Additional categories include \textit{Analytics} (performance monitoring and evaluation) and \textit{Development} activities (e.g., CI/CD pipelines and testing). Details on these categories, the complete topic hierarchy, and their distributions are available in the replication package. Since we aimed to produce as fine-grained taxonomy as possible, \textit{Caching}, due to its prevalence, was kept as a separate category instead of merging into others. We found that all categories, except \textit{Analytics}, directly focus on performance optimization. 


\begin{figure}[h]
    \centering
    \includegraphics[width=1\linewidth,keepaspectratio]{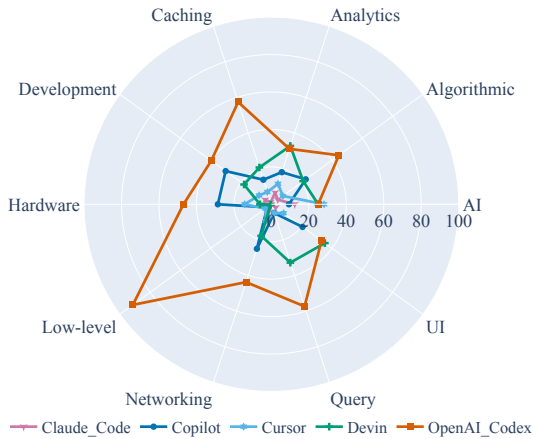}
    \caption{Percentage of PRs authored by different AI agents across categories (all PRs).}
    \Description{Percentage of PRs authored by different AI agents across categories.}
    \label{fig:agent_performance}
\end{figure}

Figure~\ref{fig:agent_performance} presents the distribution of performance-related PRs authored by different AI agents across categories. OpenAI Codex dominates most categories, especially \textit{Low-level} tasks with over 90\% of PRs. In contrast, its relative contribution is lower in the \textit{Analytics}, \textit{AI}, and \textit{UI} categories. Devin contributes its largest share of PRs in \textit{Analytics} and \textit{UI}, while Cursor contributes the most in the \textit{AI} category. Nonetheless, OpenAI Codex accounts for at least 30\% of PRs in every category. This is unsurprising, given that the majority of all PRs in the AIDev dataset~\cite{li2025rise} is dominated by OpenAI Codex. 

Overall, although we observe that certain categories appear relatively more frequently for some agents, a more diverse and balanced dataset is needed to draw definitive conclusions, representing a potential avenue for future work in the community.


\summarybox{RQ1}{We identified 52 performance-related topics, grouped into 10 categories. Overall, agentic AI systems aim to address a broad spectrum of performance optimizations across multiple layers of the software stack, ranging from \textit{Low-level} optimizations to \textit{UI}-level concerns. }




\subsection{RQ2: Acceptance Rate and Merge time}

\begin{figure}[h!]
    \centering
    \includegraphics[width=1\linewidth, keepaspectratio]{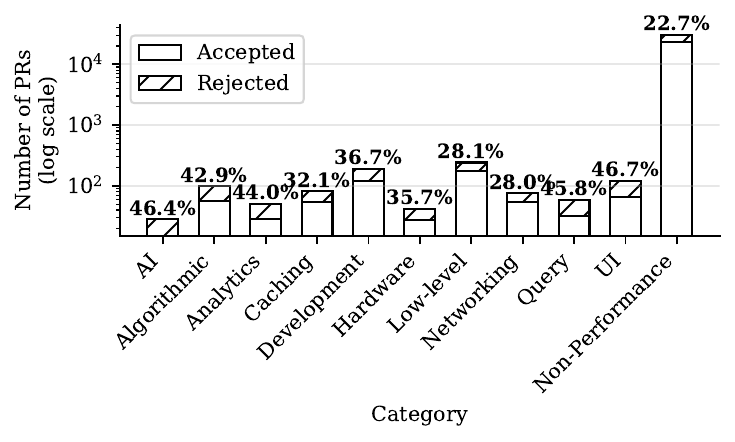}
    \caption{Number of accepted and rejected PRs. The numeric values represent the percentage of rejected PRs.}
    \Description{Number of accepted and rejected PRs for each category. The numeric values represent the percentage of rejected PRs.}
    \label{fig:accepted_rejected}
\end{figure}

Prior work reports generally low acceptance rates for agentic PRs~\cite{li2025rise}. We investigate whether this trend also applies to performance-related PRs. We consider a PR rejected if it is closed without being merged. We found that across all performance PRs, the overall rejection rate is 36.5\%, which is significantly higher than non-performance PRs with a rate of 22.7\%. The rejection rates also vary based on the type of performance optimization applied. As shown in Figure~\ref{fig:accepted_rejected}, the higher-level \textit{UI}, \textit{AI}, and \textit{Analytics} categories exhibit relatively higher rejection rates, suggesting that these performance-related changes may require more context-sensitive reasoning or human intervention. In contrast, lower-level optimizations, such as compiler-related PRs, show much higher acceptance rates.
  
\begin{figure}[H]
    \centering
    \includegraphics[width=1\linewidth,keepaspectratio]{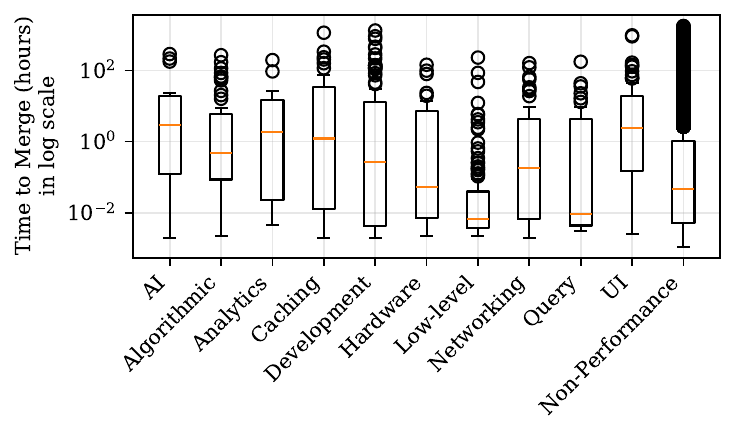}
    \caption{Time to merge (hours, log scale) across categories.}
    \Description{Time to merge (hours, log scale) across categories.}
    \label{fig:mergetime}
\end{figure}

Next, we select 625 accepted PRs and analyze their time to merge. Figure~\ref{fig:mergetime} shows that \textit{UI}, \textit{AI}, and \textit{Analytics} related PRs have the longest median merge times, whereas \textit{Low-level} PRs are merged the fastest. This pattern is very consistent with Figure~\ref{fig:accepted_rejected}. While evaluated with the Wilcoxon rank sum test, we found that the merge time distributions of the mentioned higher-level categories are statistically significantly different than the distribution in the \textit{Low-level} category. Also, according to Cliff's delta, these differences are non-negligible with large effect sizes. 

\begin{figure}[h]
    \centering
    \includegraphics[width=1\linewidth, keepaspectratio]{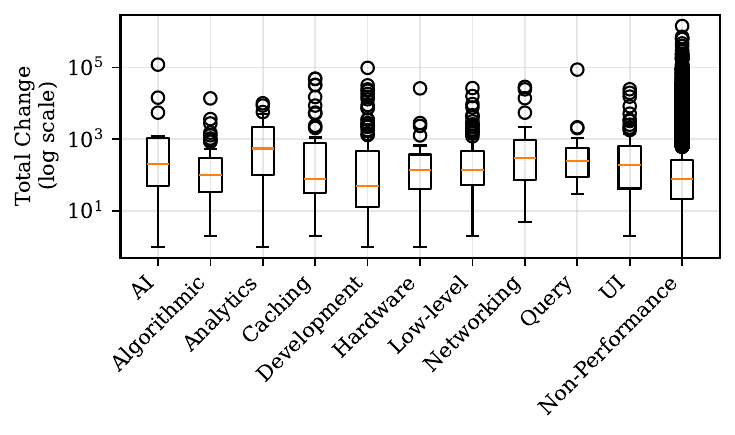}
    \caption{Total line changes (log scale) across categories.}
    \Description{Total line changes (log scale) across categories.}
    \label{fig:linechange}
\end{figure}

We can argue that it is probably not the change type, but the change size that impacts rejection rate and merge time. 
However, as shown in Figure~\ref{fig:linechange}, the distributions of the total number of code line changes (addition + deletion) are not as different as we found in merge time distributions. For example, the differences between the high-level and low-level categories are now statistically insignificant. In fact, the highest Kendall’s $\tau$ correlation between merge time and total line changes is only $0.38$ (for the \textit{Hardware} category), indicating only a weak association.

\summarybox{RQ2}
{Agentic PRs related to performance are rejected significantly more than non-performance PRs. Also, the rejection rates and review times vary significantly based on the types of optimizations applied by the agents.}

\subsection{RQ3: Association with SDLC Activities}



The AIDev-POP dataset provides PR activity type annotations across nine activities (e.g., \textit{feat}, \textit{fix}, \textit{refactor}), including a dedicated \textit{perf} type activity containing 340 PRs. However, performance-related activity is not mutually exclusive with others, as we explained in our methodology. We therefore excluded the 340 PRs labeled as \textit{perf}, as these PRs do not specify other underlying SDLC activities.

\begin{figure}[h!]
    \centering
    \includegraphics[width=1\linewidth, keepaspectratio]{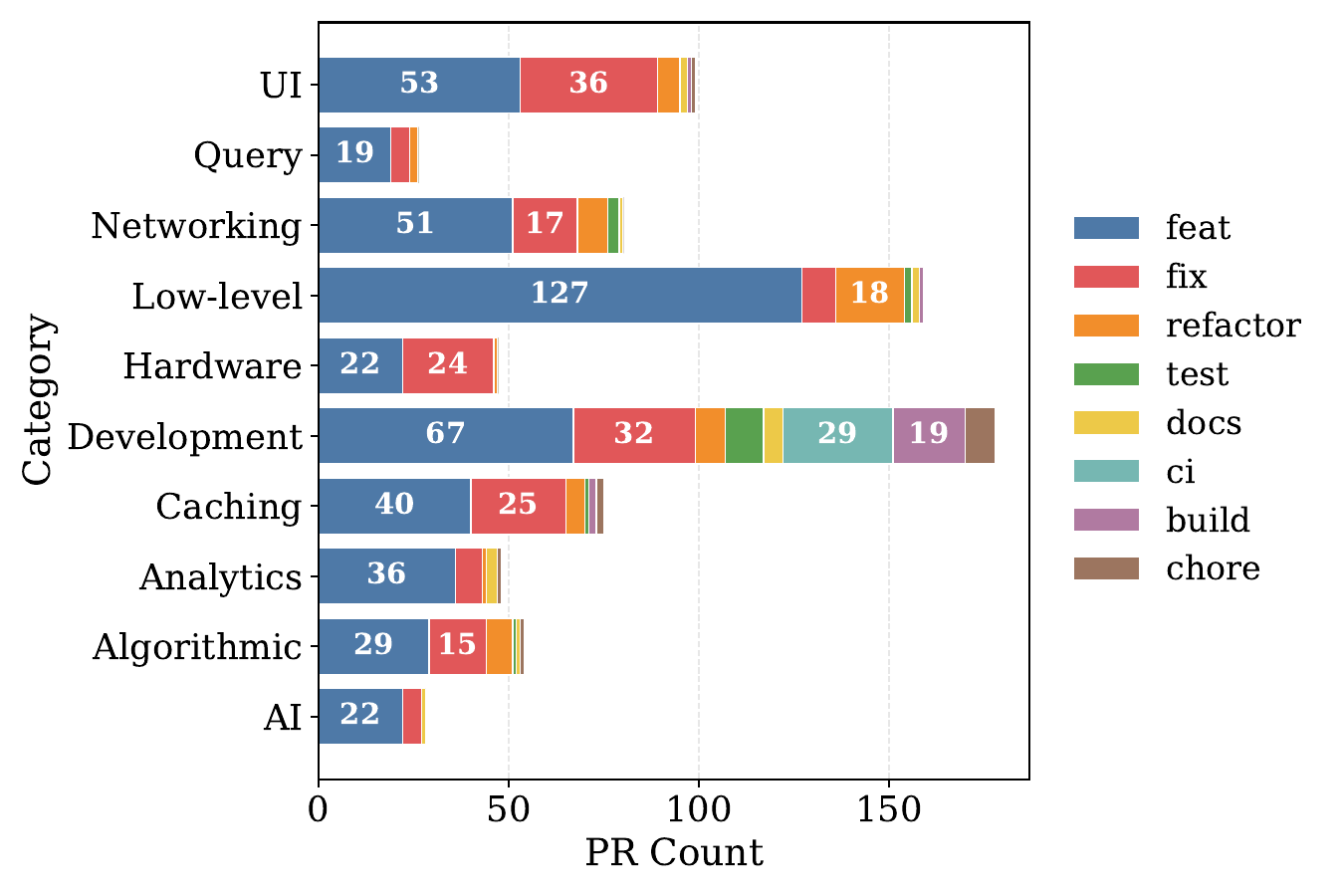}
    \caption{Distribution of PR types across categories.}
    \Description{Distribution of activity types across performance-related categories.}
    \label{fig:pr_types}
\end{figure}

Figure~\ref{fig:pr_types} shows the association of different SDLC activities with our ten identified performance-related categories. For all categories except \textit{Hardware}, most performance optimizations occur during feature implementation. While bug-fixing and refactoring activities are also observed, our results indicate that agentic AI primarily addresses performance concerns during the development phase, with comparatively less focus during testing and maintenance.

\summarybox{RQ3}{AI agents primarily address performance concerns during the development phase (e.g., feature implementation). Future agents could be trained and tuned to consistently address this critical non-functional requirement across the entire SDLC.}

\section{Threats to Validity}
\textbf{Internal Validity.}
We identified performance PRs using an LLM-based classifier, which improves upon keyword-based methods but still misclassifies ambiguous PR descriptions. This threat was mitigated due to BERTopic's ability to work on noisy data~\cite{janssens2025comparative}. Our topic labeling and categorization approach relied on manual judgment, introducing potential subjectivity despite independent annotation and consensus resolution. Differences in acceptance rates, merge times, and review effort may be affected by unobserved factors such as project contributors, review practices, or repository maturity.

\textbf{External Validity.}
Our study analyzed agentic PRs collected from a specific dataset, which may impact the generalizability of our observations.

\section{Related Works}
Recent advances in LLMs have significantly influenced SE research, motivating studies that explore their use across a wide range of development activities~\cite{zhang2023survey}. Prior work has investigated LLMs for code generation, program comprehension, and automated software maintenance, highlighting their potential to support or automate developer tasks~\cite{zheng2025towards, joel2024survey, crupi2025effectiveness}. Empirical studies further demonstrate that LLMs can generate functionally correct implementations~\cite{chen2021evaluating, wang2024performance}. As a result, LLM-based systems are increasingly being examined not only as assistive tools, but also as autonomous coding agents~\cite{li2025rise}.

Due to its importance, traditional SE research has long emphasized on performance-aware software engineering~\cite{balsamo2004model, zaman2011security, zhao_platform-agnostic_2024, das2016quantitative, jin2012understanding}. Similarly, alongside functional correctness, non-functional properties of LLM-generated code have attracted growing attention~\cite{wang2024performance,coignion2024performance, yi_experimental_2025}. Performance, in particular, is a critical concern, as inefficient code can introduce runtime overheads, scalability issues, and increased operational costs~\cite{balsamo2004model}. 
Consequently, recent works focused on analyzing the performance of LLM-generated code~\cite{chen2021evaluating,wang2024performance,watanabe_use_2025}. Some studies claim that AI-generated code performs comparably to or even better than human code~\cite{coignion2024performance,wang2024performance,peng2025perfcodegen}, while other studies show that LLMs can produce performance-improving patches, although these often underperform compared to human-optimized solutions~\cite{yi_experimental_2025}. Techniques that integrate execution feedback or explanations further improve the performance of generated code or assist in fixing performance bugs~\cite{peng2025perfcodegen, sijwali_fixing_2025}. In parallel, empirical analyses of AI agents reveal they often perform well in functionality but lack in performance~\cite{li2024assessing}.

Despite mixed results, autonomous agents are solving real software engineering tasks \cite{li2025rise}. However, it remains unclear how they address real-world performance concerns and at what levels they contribute. These are the questions that we aimed to answer. 
\section{Conclusion}


This paper presents an empirical study of performance-related pull requests generated by AI agents. Using BERTopic, we identified 52 performance topics organized into 10 categories spanning diverse layers in the software stack. We found that performance-related PRs are rejected more frequently than non-performance PRs, suggesting potential trust or verification issues with AI-generated performance optimizations. Additionally, acceptance rates and review times vary by optimization type: \textit{UI}-related PRs experience the highest rejection rates, whereas low-level optimizations are more frequently accepted and merged more quickly. Future work should focus on leveraging larger and more diverse datasets to validate the identified categories. We also need to empirically evaluate whether agent-generated changes yield measurable improvements, as such evidence is currently limited. Additionally, AI-based agents should be improved to address performance issues throughout the SDLC more consistently.

\bibliographystyle{ACM-Reference-Format}
\bibliography{sample-base, references, related_work_references}

\end{document}